\documentclass[fleqn,10pt]{wlscirep}
\usepackage[utf8]{inputenc}
\usepackage[T1]{fontenc}

\usepackage{hyperref}

\begin{document}

\title{Evaluating the impact of incomplete contact tracing data on urban epidemic dynamics}

\author[1]{Min-Kyung Chae}
\author[2]{Woo-Sik Son}
\author[3,*]{Sang Hoon Lee}
\affil[1]{Department of Physics and Astronomy, Sejong University, Seoul 05006, Republic of Korea}
\affil[2]{National Institute for Mathematical Sciences, Daejeon, 34047, Republic of Korea}
\affil[3]{Department of Physics and the Research Institute of Natural Science, Gyeongsang National University, Jinju, 52828, Republic of Korea}

\affil[*]{lshlj82@gnu.ac.kr}



\begin{abstract}
Contact tracing (CT) is a frontline measure against emerging epidemics, yet in practice it is never complete. The quantitative impact of missing information---such as untraced cases or unnotified contacts---on the effectiveness of CT remains insufficiently understood. Using a stochastic agent-based model with sociodemographics from metropolitan areas in South Korea, we simulate how different forms of information loss affect epidemic spreading dynamics. We construct information-loss scenarios based on two types: infector-omission (IO), the omission of infected individuals from the tracing process, and contact-omission (CO), the omission of specific contact events even when the infected individuals themselves are identified. The sensitivity of epidemic dynamics to increasing omission rates differs markedly between the two types: IO produces substantially stronger and more abrupt changes in transmission structure and epidemic outcomes, whereas CO produces more gradual effects. Notably, CT effectiveness breaks down beyond a city-specific threshold---an IO rate of approximately 4\% in Seoul but about 10\% in less populous Busan---underscoring that CT strategies must be tailored to regional population and mobility structure. Both IO and CO scenarios also lead to an increase in the transmission network diameter as information loss grows, indicating that a small network diameter reflects effective contact tracing that limits the depth of transmission chains. Collectively, our results offer threshold estimates and practical guidance for designing robust CT systems in the real world.
\end{abstract}

\flushbottom
\maketitle
%
%
\thispagestyle{empty}


\section*{Introduction}

The frequency of major infectious disease outbreaks has increased in recent decades, with progressively shorter intervals between successive events~\cite{daszak2008global, smith2014global}. Among these, the unprecedented COVID-19 pandemic demonstrated that emerging pathogens can severely disrupt not only individual health but also entire economies, education systems, and healthcare infrastructures~\cite{naidoo2020reset}. When effective vaccines or cures are unavailable or not yet widely deployed, as in the initial period of an epidemic, governments are compelled to rely on a broad arsenal of non-pharmaceutical interventions (NPIs) to mitigate transmission~\cite{ferguson2005strategies}. First-line responses typically include school closures, teleworking, physical distancing, and contact tracing. When public health systems are overwhelmed, authorities may resort to city lockdowns---measures that are effective in reducing transmission but entail substantial socio-economic costs~\cite{hellewell2020feasibility, gilbert2020preparing}. To delay or avoid resorting to such costly measures, traditional epidemiological responses---such as case isolation, movement trajectory reconstruction, and quarantine of contacts---remain essential, particularly during the early phase of a pandemic~\cite{kucharski2020effectiveness, fraser2004factors, cai2025assessing}. Managing these interventions effectively under uncertainty is precisely where complexity science and information systems have been argued to play a critical role, ultimately helping to save lives~\cite{helbing2015saving}.

Contact tracing (CT) identifies the trajectories of confirmed cases and rapidly finds persons whose paths overlap with those trajectories for testing and quarantine. This approach is especially effective when case numbers are still low~\cite{cetron2004isolation}. However, CT is far from perfect and remains highly dependent on human resources. Investigators rely on the recollection of individuals to reconstruct trajectories, manually compile contact lists, and request tests and quarantine. These steps inevitably lead to uncertainty and incompleteness specifically, the inability to trace confirmed cases and the failure to notify contacts. To address these limitations, various complementary strategies have been explored~\cite{kojaku2021effectiveness}. For example, network-based computational approaches identify transmission sources directly from network structure and temporal dynamics~\cite{tang2025observer}, and many countries have experimented with mobile-application-based digital tracing~\cite{ferretti2020quantifying, almagor2020exploring, lopera2025impact}. Despite these efforts, full coverage remains elusive, often attributed to privacy concerns, varying levels of smartphone adoption, and low user compliance~\cite{altmann2020acceptability}.

Quantitative assessments of CT must account for the information omissions that arise during real-world implementation. However, few quantitative studies have explicitly represented such omissions within the structure of the contact network itself. We developed a high-resolution agent-based model (ABM) to evaluate how such omissions affect CT effectiveness by simulating the spread of emerging infectious diseases across metropolitan areas in South Korea. The model incorporates a multilayer contact network including households, school classrooms, workplaces, friendships, and local communities. We independently examine two types of information omission: (1) infector-omission (IO), where the movements and interactions of confirmed cases are not reconstructed; and (2) contact-omission (CO), where individuals who have been in contact with confirmed cases are not informed in time. We further distinguish CO into selective (SCO) and uniform (UCO) scenarios, depending on how omissions are distributed across contact networks.

This paper is organized as follows. First, in the ``Methods'' section, we describe a stochastic ABM based on a synthetic population that reflects the demographic and social contact structure of Seoul and Busan. 

In the ``Results'' section, we quantitatively analyze the impact of the two types of information omission (IO and CO) on the spread of infection by comparing simulation results between Seoul and Busan, two cities with contrasting age distributions. Finally, in the ``Discussion'' section, we summarize the strengths and limitations of this study and discuss policy implications for designing effective CT strategies under realistic constraints.

\section*{Methods}
This section describes our ABM developed to evaluate the effectiveness of CT in response to the emergence of an infectious disease in Seoul (Busan). The model incorporates multilayer social networks~\cite{Kivela2014} to capture heterogeneous interpersonal interactions across different social contexts. Agents carry out typical daily activities such as going to school, commuting to work, meeting friends, during which they come into contact with other agents based on realistic schedules. These interactions generate dynamic and diverse contact patterns, allowing the model to simulate the complexity of real-world transmission environments.

Infectious diseases are transmitted through contact between individuals. Therefore, understanding transmission dynamics requires constructing social network models that reflect the underlying interpersonal interaction structure within the population. Given the importance of contact patterns especially in the context of CT the ABM is preferable to traditional compartmental models due to its ability to incorporate individual-level heterogeneity~\cite{de2014agent, rahmandad2004heterogeneity, markovivc2021socio, sunahara2023complexity}. Beyond traditional compartmental models, this individual-level resolution allows our ABM to explicitly reconstruct the directed transmission network. This capability enables us to monitor structural properties such as network diameter, which serves as a proxy for the number of transmission generations, offering deeper insights into spreading dynamics beyond simple prevalence. Network-based approaches have more broadly provided valuable insights into spreading dynamics, including epidemic processes on multilayer networks~\cite{wang2025epidemic} and the spatial invasion of infections across metapopulation networks~\cite{wang2018uncovering, wang2015identifying}.

We have created a synthetic population reflecting the sociodemographic characteristics of Seoul's (Busan's) individuals, which enables more realistic simulations. The model explicitly tracks interactions at the individual level, allowing interpersonal contacts to be distinguished over time.

\subsection*{Agents} 
We construct the synthetic population to represent the population of Seoul (Busan), matching the population size and incorporating details such as household ID, age, and residence. This is based on the 2020 Korean census data~\cite{2020census} (also used in Chae \emph{et al}.~\cite{chae2023evaluation}), which sampled 2\% of the population and was released by the MicroData Integrated Service. We project the 2\% census of Seoul (145,817 records) to the entire population of Seoul (9,529,266 agents) using an iterative proportional updating algorithm~\cite{ye2009methodology, chae2023evaluation}. We also project Busan (61,217 records $\rightarrow$ 3,313,542 agents).

Figure~\ref{fig:abm}(a) illustrates the agent attributes. Each agent is characterized by sociodemographic attributes (age and residence), social affiliations across the five contact-network layers (households, school classrooms, workplaces, friendships, and local communities), and an epidemiological state (susceptible, exposed, infectious, or recovered). During the implementation of CT policies, a quarantine status is also assigned. The spatial structure of the model reflects the administrative level-2 (ADM-2) regions.

\subsection*{Social Networks}
\begin{figure}
\centering
\includegraphics[width=\textwidth]{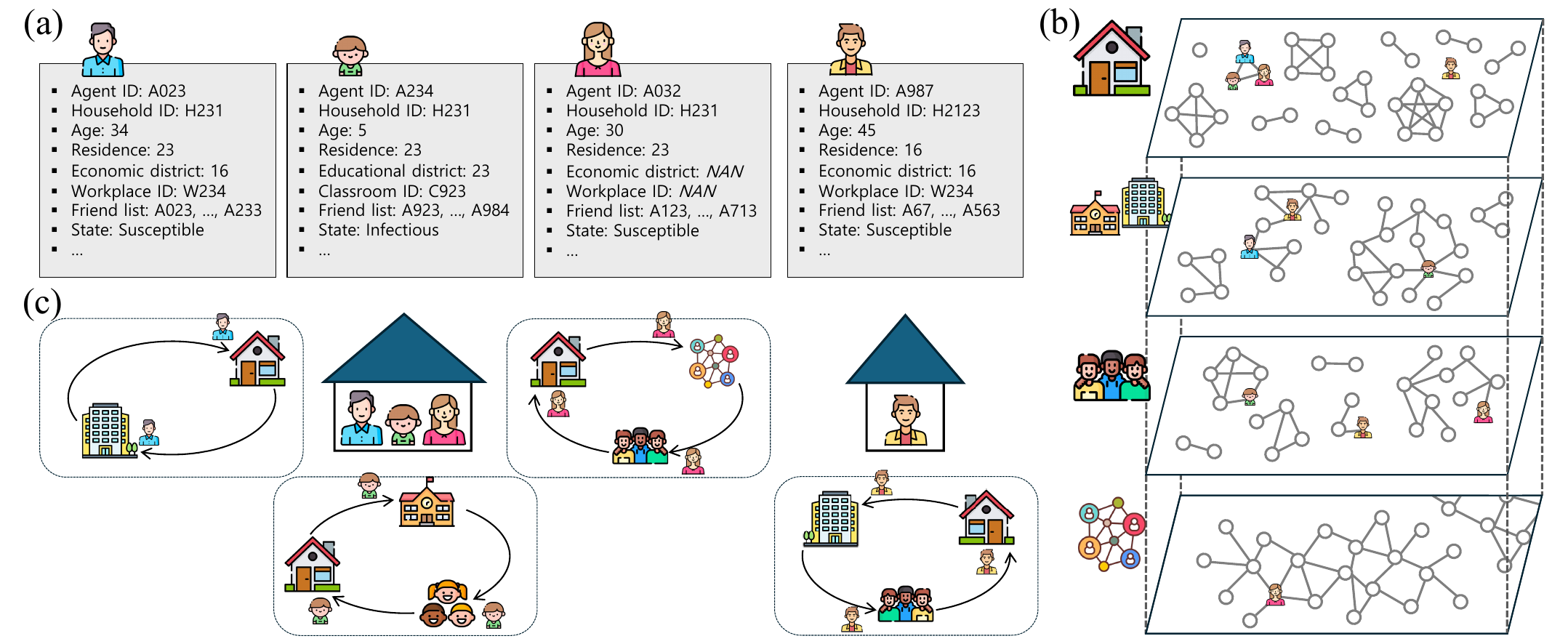}
\caption{Synthetic population and multilayer contact structure. (a) Schematic representation of each agent's sociodemographic attributes, including household ID, age, residence, workplace and school classroom affiliation, and epidemiological status. (b, c) Example of a contact network generated by daily behavior patterns over a single day. (b) Multilayer contact network comprising household, workplace/classroom, friendship, and local-community interactions. Each layer captures the social connection among agents. (c) Example of a daily routine for each agent. Agents follow different schedules depending on their demographic characteristics, such as attending school, going to work, or interacting with friends and community members. The agent icons were sourced from FLATICON~\cite{FLATICON}.}
\label{fig:abm}
\end{figure}

Figure~\ref{fig:abm}(b) shows the multilayer contact network consisting of five social structures:

\begin{enumerate}
    \item \textbf{Households}: Agents sharing the same household ID are assigned to the same housing unit within an ADM-2 region. All agents have a household ID.
    
    \item \textbf{Workplaces}: We use the regional employment rate~\cite{eco_rate} to randomly select working-age agents (ages 19--64) from the synthetic population and designate them as workers. Each selected agent is assigned a workplace ID and an economic district based on commuting data. Workplaces are generated as discrete sites ranging from small offices (mean size = 5 agents) to large offices (mean size = 10 agents), reflecting the distribution of workplace sizes in Seoul (Busan)~\cite{workplacesize}. Each worker is associated with a single workplace.
    
    \item \textbf{School classrooms}: Students are created in much the same way as workers. Based on the regional attendance rates~\cite{kindergarten, elementary, junior, high}, we randomly select school-age agents (ages 3--18) from the synthetic population and designate them as students. Each student's classroom ID and educational district are assigned using commuting data. Classroom size is determined by the average number of students per class in Seoul (Busan)~\cite{kindergarten, elementary, junior, high}. The number of teachers in each classroom is set using the average number of teachers per student, so that it scales with the classroom's student count. All classmates are of the same age. School institutions are not explicitly modeled; classrooms are represented as isolated units, so that students from different classrooms within the same school do not interact.

    \item \textbf{Friendships}: A friendship network is generated using a homophilic Barab{\'a}si-Albert (BA) model~\cite{lee2019homophily, barabasi1999emergence} that preferentially connects agents of the same age, reflecting the cultural importance of age in Korean social relationships. Agents are first organized into age cohorts, each spanning a decade (ages 0--9, 10--19, \dots, 80 and older). Within each cohort, the homophilic BA network is generated in blocks of 1,000 agents, where new agents form preferential attachments to existing ones and a homophily parameter $h$ assigns a high linking probability ($h = 0.9$) to pairs of agents of the same age and a low probability ($1 - h = 0.1$) to pairs of different ages. Consequently, connections are formed only within a block, with no links established across blocks or cohorts. This process is applied both within and across administrative regions, resulting in an undirected network. Each agent has an average of 20 friends ($\sigma$ = 13), ranging from 7 to 228.    
    \item \textbf{Local communities}: Agents living in the same district are connected to simulate casual, short-term encounters in the local community. The connections in the local-community network are not static.
\end{enumerate}

Agents follow daily routines according to their demographic profiles (Figure~\ref{fig:abm}(c)). All agents interact with their household members daily, and on weekdays students attend their classrooms while workers go to their workplaces. Contacts within the friendship and local-community layers, by contrast, are modeled stochastically: each day, every agent independently decides whether to meet friends with a probability of 1/7, and an interaction occurs only when both agents choose to do so. As a result, the structure of the friendship network varies from day to day. The local-community layer follows the same mechanism, except that local-community interactions may involve encounters with unfamiliar individuals as well as acquaintances. Infectious disease spreads through these social interactions across all layers. Further details of the network construction process are provided in the Supplementary Information.

\subsection*{Epidemiological Parameters}
\begin{figure}
\centering
\includegraphics[width=12.0cm]{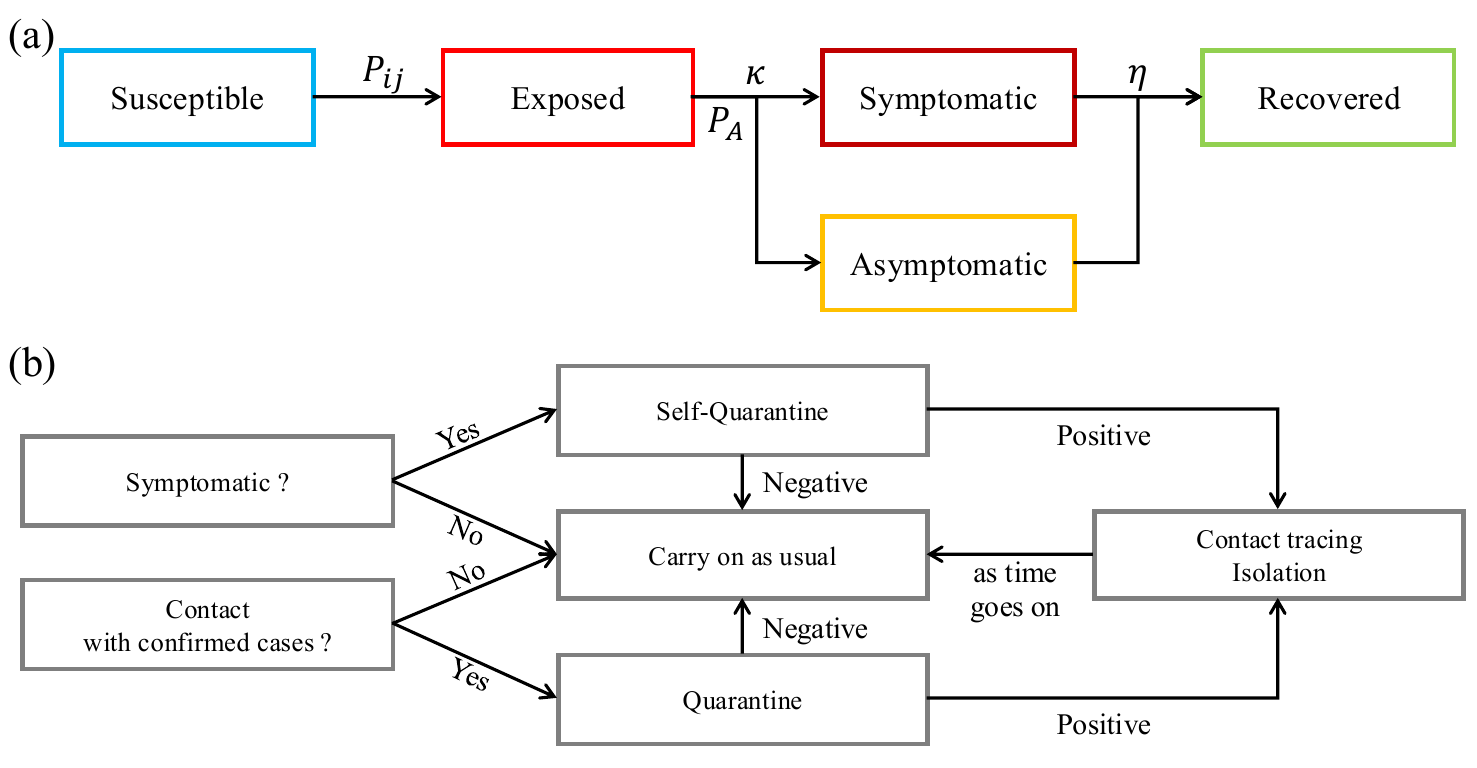}
\caption{Disease progression and contact-tracing protocol. (a) Progression of an infectious disease through five states: susceptible ($S$), exposed ($E$), symptomatic infectious ($I_S$), asymptomatic infectious ($I_A$), and recovered ($R$). A susceptible agent becomes exposed with transmission probability $P_{ij}$ upon contact with an infector ($S \rightarrow E$). After $\kappa$ days in the exposed state, agents transition to either $I_S$ or $I_A$ ($P_A$), and recover after $\eta$ days. (b) Schematic representation of the CT protocol, illustrating the pathways of symptomatic agents and exposed contacts through testing, (self-)quarantine, isolation, and secondary contact identification.}
\label{fig:seir}
\end{figure}

We model disease progression using an extended susceptible-exposed-infectious-recovered (SEIR) model described in Figure~\ref{fig:seir}(a). All agents are initially susceptible ($S$), except for a small number of initially exposed ($E$) agents. Upon contact with an infector $j$, a susceptible agent $i$ becomes exposed ($E$) with a probability of transmission $P_{ij}$. The exposed period ($\kappa$) is drawn from a gamma distribution $f(\kappa;\alpha,\theta)=\frac{1}{\Gamma(\alpha)\theta^{\alpha}}\kappa^{\alpha-1}e^{-\kappa/\theta}$ (shape $\alpha$ = 1.926, scale $\theta$ = 1.775) for each agent~\cite{he2020temporal}. Agents in this state are initially non-infectious but become infectious up to two days before entering the infectious state ($I$)~\cite{sun2021transmission}. Infectiousness is modeled as the product of viral shedding and relative infectiousness. Viral shedding follows a gamma distribution with a mean of 3.067 and a standard deviation of 2.109~\cite{he2020temporal}. Relative infectiousness ($\xi_j$) varies across agents, such that some agents exhibit higher or lower overall infectiousness levels. After the exposed state, agents transition to either asymptomatic or symptomatic infectious ($I_A$ or $I_S$) states with a 20\% probability of being asymptomatic ($P_A = 0.2$). All infectious agents are assumed to recover after 8 days ($\eta = 8$)~\cite{wolfel2020virological}, after which they are no longer infectious. Reinfection is not considered in the study, so that recovered agents gain permanent immunity.

At the end of each simulation day, infection events are determined stochastically, with a time step $\rm{d}t$ of 1 day. Each $S$ agent $i$ computes the infection risk independently for all contacts with $E$ and $I$ agents $j$ during the day. The probability of transmission $P_{ij}$ per contact is defined as:

\begin{equation}
P_{ij} = 1 - e^{-\lambda_{ij}}, \quad \lambda_{ij} = t^{ij}_{n} \varphi_j.
\label{eq:transmission_probability}
\end{equation}

Here, $\lambda_{ij}$ represents the force of infection associated with a single contact, $t^{ij}_{n}$ is the contact duration by location obtained from South Korean social close contact survey~\cite{son2025social}, and $\varphi_{j}$ is the infectiousness of the infector $j$. The distributions of contact duration by location are described in the Supplementary Information (SI). Because empirically observed contact durations already reflect relative contact intensities across locations, no additional network-specific scaling factor is introduced in $\lambda_{ij}$. To implement this transmission mechanism, we apply a Monte Carlo sampling procedure based on an accept--reject scheme: for each contact, a uniform random number is drawn from the interval $[0,1]$, and infection occurs if this value is less than $P_{ij}$. This procedure preserves the heterogeneity in contact duration and infectiousness. The specific numerical values of the epidemiological parameters are summarized in Table~\ref{tab:model_parameter}, and additional details are provided in the SI.

\begin{table}
\centering
\caption{Summary of parameters and assumptions used in the simulation of CT. The probability density function of the gamma distribution is $f(x;\alpha,\theta)=\frac{1}{\Gamma(\alpha)\theta^{\alpha}}x^{\alpha-1}e^{-x/\theta}$.}

\begin{tabular}{|l|c|c|p{6.3cm}|}
\hline
\textbf{Parameter} & \textbf{Symbol} & \textbf{Value / Distribution} & \textbf{Description} \\
\hline
simulation time step & $\Delta t$ & 1 day & time unit of simulation \\
\hline
friend contact probability & --- & $1/7$ per day & daily chance of friends' gathering \\
\hline
local contact probability & --- & $1/7$ per day & daily chance of local random contact \\
\hline
exposed period & $\kappa$ & $f(x; 1.926, 1.775)$~\cite{he2020temporal} & period from exposure $E$ to infectious $I$\\
\hline
infectious period & $\eta$ & 8 days~\cite{wolfel2020virological} & period from infectious $I$ to recovered $R$ \\
\hline
asymptomatic probability & $P_A$ & 0.2~\cite{sow2026generalized}& fraction of asymptomatic infectious cases \\
\hline
transmission probability & $P_{ij}$ & $1 - e^{-\lambda_{ij}}$ & probability of infection given contact \\
\hline
force of infection & $\lambda_{ij}$ & $ t^{ij}_n \times \varphi_j$ & infection risk per contact \\
\hline
contact duration & $t^{ij}_n$ & survey-based & average contact duration per layer, derived from South Korea’s social contact survey~\cite{son2025social} \\
\hline
infectiousness & $\varphi_j$ & $\text{viral shedding} \times \xi_j$~\cite{ferguson2005strategies} & viral shedding adjusted by individual infectiousness \\
\hline
viral shedding & --- & $f(x; 3.067, 2.109)$~\cite{he2020temporal} & baseline viral shedding level of the disease \\
\hline
relative infectiousness & $\xi_j$ & $f(x; 1, 0.5)$~\cite{ferguson2005strategies} & individual infectiousness of infector $j$ \\
\hline
self-testing probability & $P_t$ & 0.5 & probability that symptomatic agents seek testing voluntarily \\
\hline
quarantine duration & --- & 1 day & period of quarantine before result \\
\hline
self-quarantine duration & --- & 1 day & period of voluntary quarantine before result \\
\hline
isolation duration & --- & 7 days & period of isolation after testing positive \\
\hline
initial exposed agents & $E_0$ & 20 (40) & number of initial $E$ population \\
\hline
\end{tabular}
\label{tab:model_parameter}
\end{table}

We summarize in Table~\ref{tab:model_parameter} the simulation parameters and behavioral rules implemented in our ABM. These values are applied consistently across all simulations unless otherwise specified. Because this study assumes the emergence of a novel infectious disease rather than a specific known pathogen, most epidemiological parameters are set to assumed values, whereas those for which empirical estimates are available, such as the exposed period and the infectious period, are adopted from COVID-19 studies~\cite{he2020temporal, sun2021transmission, wolfel2020virological}. When a new infectious disease emerges in the future, the model can be readily adapted by updating these epidemiological parameters.

\subsection*{Contact Tracing}
Under the CT policy, epidemiological investigators reconstruct the past movement trajectories of confirmed cases (individuals who have tested positive) to identify contacts---people who shared a space-time overlap with the case---and request that these contacts undergo testing and quarantine.

Figure~\ref{fig:seir}(b) illustrates the decision process of CT, including testing, (self-)quarantine, isolation, and recursive tracing of secondary cases. Symptomatic individuals may initiate self-quarantine before official notification, and in this model 50\% of symptomatic agents are assumed to voluntarily seek testing ($P_t = 0.5$). If a symptomatic agent officially tests positive, it is isolated and CT is conducted; if the result is negative, the self-quarantine is lifted. Contacts identified by investigators as having crossed paths with a confirmed case are tested and placed under precautionary quarantine until their results are available. Those who test negative resume their daily activities, whereas those who test positive are isolated, thereby initiating a new round of trajectory checks. Agents are also tested upon release from quarantine and re-isolated if they test positive. This dynamic and partially voluntary process captures the recursive nature of CT in real-world implementation and allows its effectiveness to be evaluated under varying conditions.

\begin{figure}
\centering
\includegraphics[width=\textwidth]{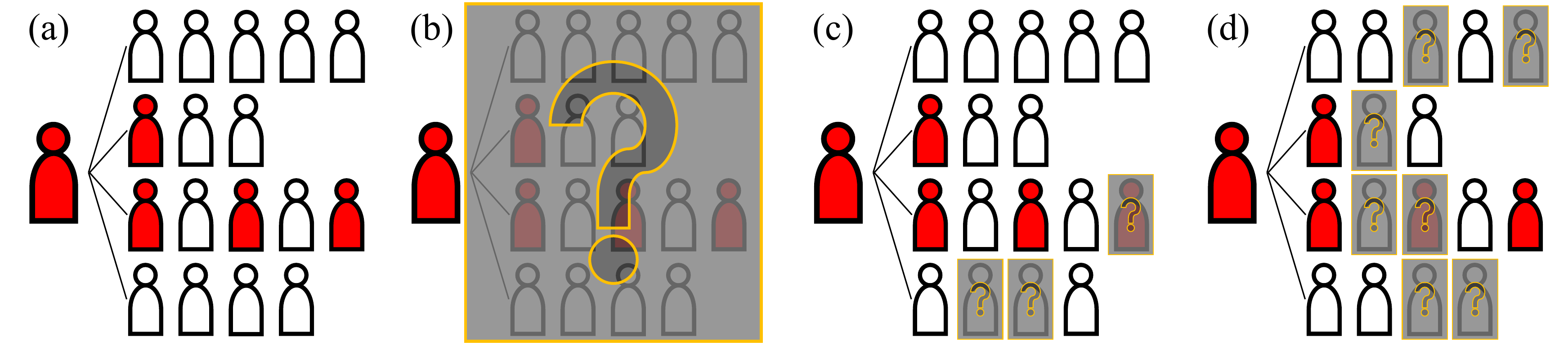}
\caption{Illustration of information-gap scenarios in manual CT. The red agents represent infected individuals, the white agents represent uninfected agents, and the agents shaded in gray with the question mark (regardless of colors inside) indicate omitted agents. (a) Ideal CT with full trajectory tracing and complete identification of contacts. (b) Infector-omission (IO): the confirmed case’s trajectory is not traced, leaving all contacts unidentified. (c) Selective contact-omission (SCO): only contacts in friends' gathering and local-community networks are partially missed. (d) Uniform contact-omission (UCO): omissions occur across all networks. 
}
\label{fig:omission}
\end{figure}

In manual CT, this entire procedure is carried out largely by people, so it depends heavily on the memory and cooperation of cases and is prone to error. As the number of cases grows, the investigators' workload intensifies, making manual CT both resource-intensive and error-prone. In practice, these constraints inevitably result in information omissions. We consider three scenarios involving two types of information omission in manual CT (Figure~\ref{fig:omission}).
\begin{itemize}
\item[$\bullet$] (Scenario 1, IO) An \textbf{infector-omission} (IO) scenario occurs when a confirmed case tests positive but their movement trajectory is not traced at all (Figure~\ref{fig:omission}(b)). In this situation, although the case is isolated, all contacts associated with that case remain unidentified and unquarantined.
\item[$\bullet$] (Scenario 2, CO) The second type, \textbf{contact-omission} (CO), occurs when agents who have been in contact with confirmed cases are not tested or quarantined. The trajectory of the infector may be available, but investigators fail to notify all agents who overlapped in time and space with the case's trajectory. This results in potentially infected agents moving freely within society, further contributing to disease spread. To capture different patterns of omission, we consider two distinct contact-omission scenarios:
\begin{itemize}
\item[$\bullet$] (Scenario 2-1, SCO) A \textbf{selective contact-omission} (SCO) scenario occurs when omissions take place only in the friendship and local-community networks (Figure~\ref{fig:omission}(c)).
\item[$\bullet$] (Scenario 2-2, UCO) A \textbf{uniform contact-omission} (UCO) scenario occurs when omission rates are applied equally across all social networks except the local-community network with the fixed omission rate 50\% (Figure~\ref{fig:omission}(d)).
\end{itemize}
\end{itemize}
In all scenarios, the omission rate for the local-community network is fixed at 50\%, reflecting the relatively high uncertainty of CT in public spaces. Each scenario is assumed to occur independently, and simultaneous occurrences are not considered.

In our simulations, these three scenarios of information omission are modeled to examine how they affect the effectiveness of CT. Since manual CT is an essential policy in the early stage of an emerging infectious disease outbreak, we assume that the number of initial cases is small: the initial exposed population is set to 20 agents ($E_0 = 20$), each remaining in the exposed ($E$) state on the first simulation day. These 20 agents are randomly selected from the synthetic population and fixed across all runs. When one of the initially exposed agents develops symptoms, visits a hospital, and tests positive, the CT policy is initiated. Each simulation is repeated over 100 independent runs to ensure statistical robustness. At the start of the simulation, there are no infectious ($I$) or recovered ($R$) agents; all agents other than the initial exposed ones are susceptible ($S$). We simulate the outbreak in virtual Seoul and Busan until it completely disappears. By varying the proportion of missing information in each scenario, we estimate the range of omission rates under which CT can still effectively mitigate the spread of infection despite being imperfect. Results for an alternative initial condition with 40 initially exposed agents are provided in the SI.

\section*{Results}

\subsection*{Impact of infector-omission on manual CT effectiveness}
The IO scenario captures situations in which some confirmed cases are not traced, leaving their movement trajectories unreconstructed and their contacts unnotified. Although untraced cases are isolated upon confirmation, their contacts continue normal activities, allowing hidden transmission chains to persist. To quantify how IO weakens containment through CT, we simulate outbreaks under varying IO rates, defined as the proportion of confirmed cases whose movement trajectories are not reconstructed by investigators. This setup enables us to evaluate how IO amplifies epidemic spread and to assess the robustness of manual CT under real-world constraints.

\begin{figure}
\centering
\includegraphics[width=0.6\textwidth]{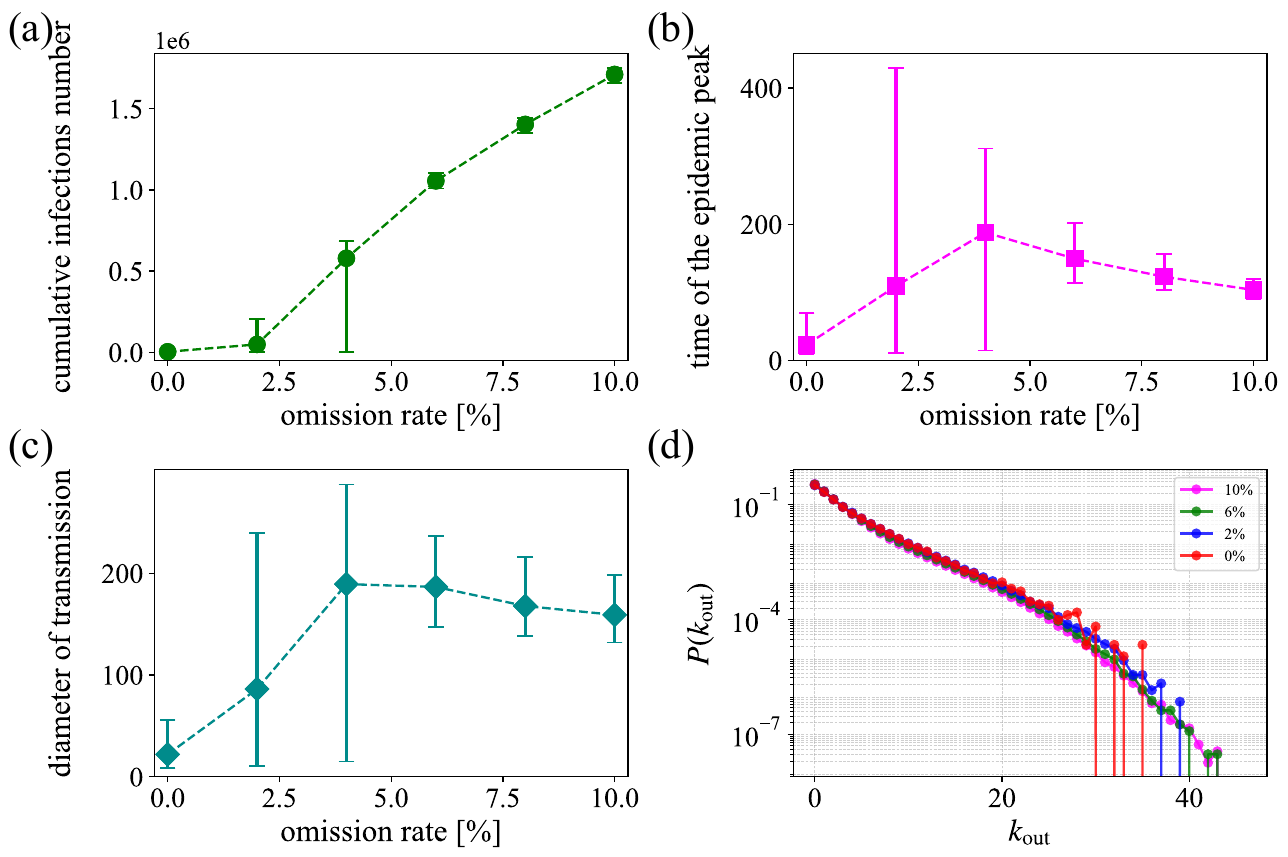}
\caption{Simulation results of the spread of infection by IO rates in virtual Seoul. Results are shown for $E_0 = 20$ with a 95\% confidence interval (CI). (a) The mean cumulative number of infections, (b) the mean time of the epidemic peak, (c) the mean diameter of the directed transmission network, and (d) the out-degree distribution of the directed transmission network (with the log-scale vertical axis). The colors represent IO rates of 0\% (red), 2\% (blue), 4\% (green), and 10\% (magenta), respectively.}
\label{fig:tor_result}
\end{figure}

We first examine how varying the IO rate affects epidemic magnitude and timing. Figure~\ref{fig:tor_result} summarizes the epidemic dynamics in virtual Seoul under varying IO rates. As shown in Fig.~\ref{fig:tor_result}(a), the mean cumulative number of infections increases sharply once the IO rate exceeds 2\%. Figure~\ref{fig:tor_result}(b) shows the mean epidemic peak time, defined as the time at which the daily number of $E$ agents reaches its maximum. Ideally, aggressive contact tracing leads to rapid containment, causing the epidemic to die out quickly (as shown in the 0\% omission rate). As the IO rate increases from 0\% to 4\%, the system undergoes a transition from a containment phase to an outbreak phase. In this range, the epidemic duration lengthens, and the peak time is delayed not caused by successful mitigation, but rather by persistent infection chains that avoid early stochastic extinction. Once the IO rate exceeds the threshold of 4\%, the dynamics shift to a typical epidemic spread where higher transmission potential leads to an earlier and higher peak.

To further examine the structural characteristics of transmission, we analyze the epidemic as a directed transmission network. Each node in the network represents the agent, and a directed edge is drawn from the infector to the infectee, generating the directed transmission network. As the IO rate increases, the mean diameter of the directed transmission network expands significantly before saturating (Figure~\ref{fig:tor_result}(c)). Here, the diameter of the network refers to the longest chain of infection, that is, the maximum number of agent-to-agent transmission steps in the network. A larger diameter indicates that the infection chains are penetrating deeper into the population. This `deepening' confirms that the containment policy fails to sever transmission links at an early stage, allowing the outbreak to sustain itself through long, uninterrupted lineages. Figure~\ref{fig:tor_result}(d) shows the out-degree distribution of the directed transmission network (plotted in the semi-log scale with the log-scale vertical axis). The network denotes the order in which infections occur, with each directed edge indicating a transmission from an infected individual to another individual. This distribution corresponds to the case reproduction number in epidemics, which is defined as the number of secondary infections generated by a single infected individual. In other words, even as the IO rate increases, the nature of the distribution remains largely unchanged. This pattern in the directed transmission network arises because the number of individuals one can meet within a limited time period is structurally constrained.

\begin{figure}
\centering
\includegraphics[width=\textwidth]{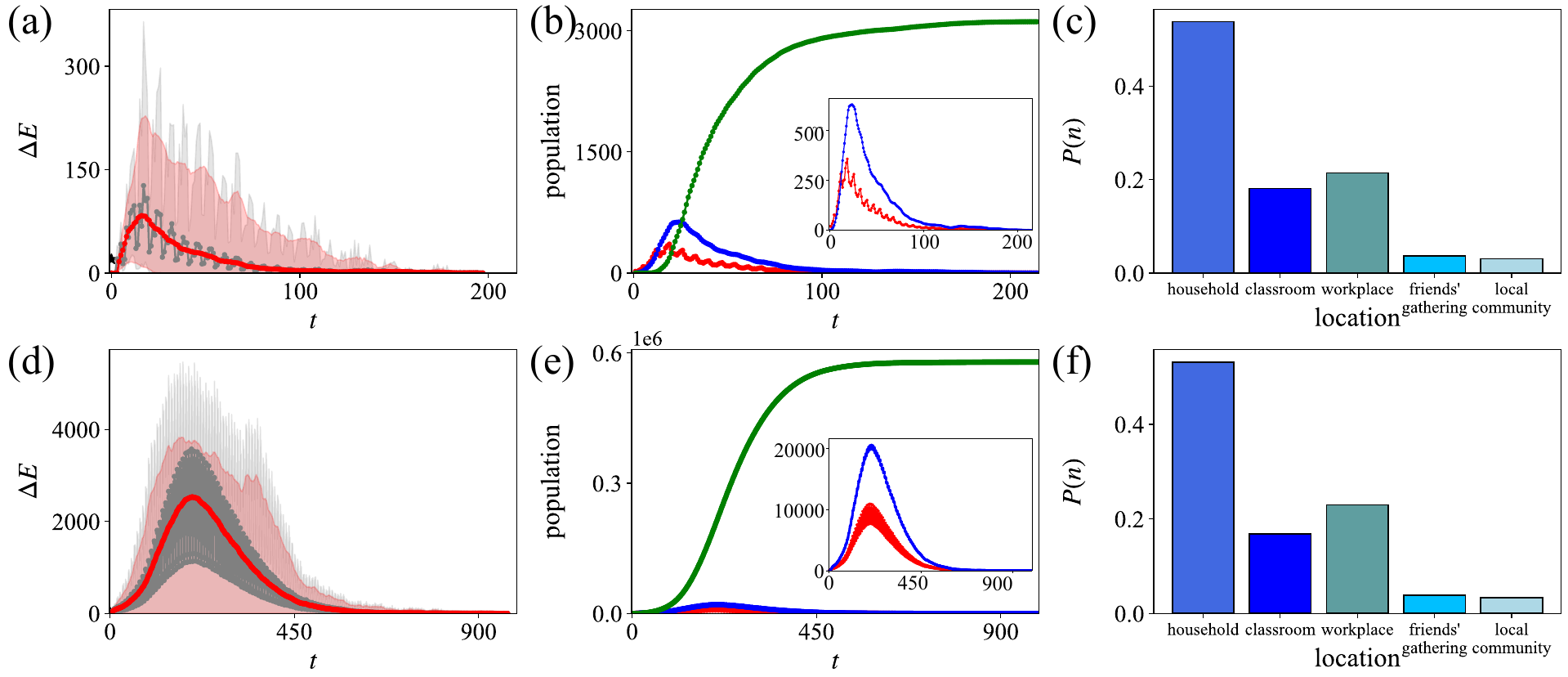}
\caption{Simulation results of the spread of infection in virtual Seoul below and above the IO threshold, with the initial number of confirmed cases set to 20. The IO rates are (a--c) 0\% and (d--f) 4\%. (a, d) The daily number of newly exposed agents, $\Delta E$, with a 95\% CI shown as shaded areas. The gray line is the daily incidence of infection, and the red line results from a 7-day moving average on the daily incidence of infection. The black star symbol denotes the initial exposed population ($E_0 = 20$). (b, e) Prevalence of $E$ (red), $I$ (blue), and $R$ (green) populations. (c, f) Distribution of probabilities, $P(n)$, for the types of locations n in which each infector causes secondary transmissions.}
\label{fig:tor_result_0_4}
\end{figure}

In Figure~\ref{fig:tor_result_0_4}, we compare epidemic dynamics at an IO rate of 4\% with those at 0\%. At an IO rate of 0\%, infections die out more rapidly, whereas a higher IO rate leads to prolonged transmission. This difference is reflected in the daily incidence trajectories, which show sustained infection at an IO rate of 4\% compared to a rapid decline at an IO rate of 0\% (Figure~\ref{fig:tor_result_0_4}(a,d)). The corresponding prevalence of the $E$, $I$, and $R$ populations further illustrates the persistence of infection under higher IO rates (Figure~\ref{fig:tor_result_0_4}(b,e)). Figure~\ref{fig:tor_result_0_4}(c, f) represents the distribution across social network layers (multilayers) where each infector generates secondary cases. Since the infection spread model has a duration-based transmission probability, most secondary infections occur within households, followed by workplaces, school classrooms, and friend gatherings. Consistent with prior evidence that household transmission constitutes a major component of respiratory disease spread~\cite{madewell2020household}.These results demonstrate that our model can realistically produce infection spread patterns under conditions where manual CT is incomplete or partially fails to identify some of the contacts and their transmission pathways.

We then extend the analysis to virtual Busan. Busan has a population of about 1/3 of Seoul, and its age distribution is different~\cite{kim2021exploratory}. It has a larger elderly population than Seoul, so the number of workers and students is relatively smaller than in Seoul. In addition, the proportion of workers and students who commute to other areas is about 4.5\%pt lower (see, SI).

\begin{figure}
\centering
\includegraphics[width=0.6\textwidth]{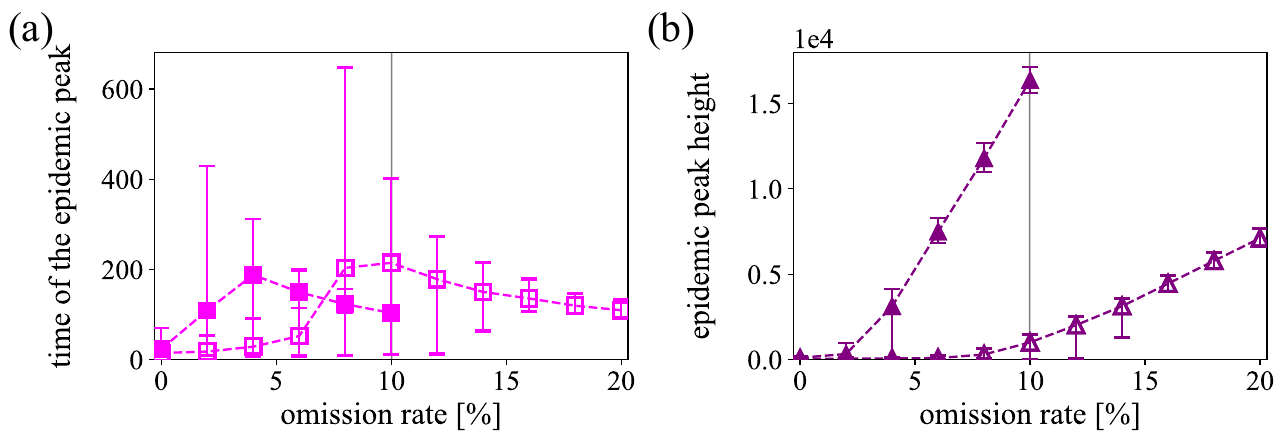}
\caption{Comparison of infector-omission effects on epidemic spread in virtual Seoul and Busan. Results are shown for $E_0 = 20$ with a 95\% CI: (a) the mean time of the epidemic peak and (b) the mean epidemic peak height against the IO rates. Filled symbols denote Seoul and open symbols denote Busan. The vertical gray line at an IO rate of 10\% is a visual guide.}
\label{fig:tor_diff}
\end{figure}

The simulation results in virtual Busan are similar to those in virtual Seoul, despite differences in population structure and demographic characteristics. Figure~\ref{fig:tor_diff} compares the two cities: the top panel shows results for virtual Seoul, and the bottom panel shows results for virtual Busan. Figure~\ref{fig:tor_diff}(a, c) show the mean time of the epidemic peak, and Figure~\ref{fig:tor_diff}(b, d) show the mean epidemic peak height. The thresholds for the IO rate are 4\% in virtual Seoul and 10\% in virtual Busan. We attribute this to the three-fold difference in population size. Despite differences in the precise threshold values due to demographic and population factors, both cities show consistent overall trends in how IO rates affect epidemic dynamics. We examine the effects of IO. We next explain the impact of CO, which represents another major source of information loss in manual CT.

To confirm that these effects are not due to the initial conditions, we double the initial exposed population ($E_0 = 40$) and repeat the same simulations. The results show the same infection spread trend as when $E_0 = 20$ (see, SI).

\subsection*{Impact of contact-omission on manual CT effectiveness}
The CT also includes requiring testing and quarantine for people whose movements overlap (contacts). However, if the number of confirmed cases becomes too large or staffing is insufficient, investigators may be unable to notify contacts about testing and quarantine in a timely manner. We simulate outbreak dynamics under CO scenarios where contacts are omitted from the notification list. We define the CO rate as the proportion of contacts who fail to receive a testing-and-quarantine notice due to tracing limitations. Two CO scenarios are constructed: (2-1) the SCO scenario, in which only contacts encountered through friend gatherings and local community interactions fail to receive notifications, and (2-2) the UCO scenario, in which contacts across all social networks may fail to be notified. In both scenarios, however, the CO rate for the local community network is fixed at 50\%, reflecting the relatively high uncertainty in tracing contacts in public spaces.

\begin{figure}
\centering
\includegraphics[width=0.6\textwidth]{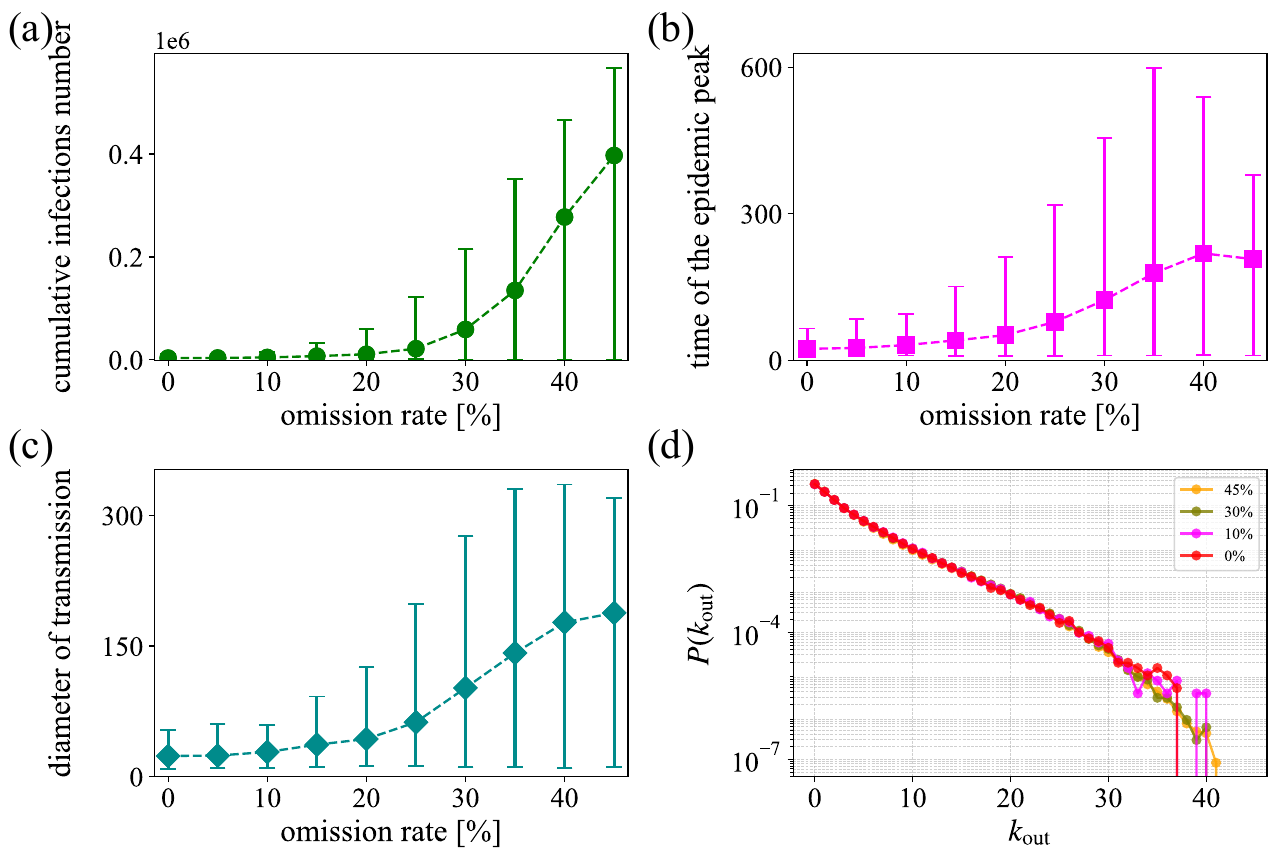}
\caption{Simulation results of the spread of infection by SCO rates in virtual Seoul. Results are shown for $E_0 = 20$ with a 95\% CI. (a) The mean cumulative number of infections, (b) the mean time of the epidemic peak, (c) the mean diameter of the directed transmission network, and (d) the out-degree distribution of the directed transmission network (with the log-scale vertical axis). The colors represent SCO rates of 0\% (red), 10\% (blue), 30\% (green), and 45\% (magenta), respectively.}
\label{fig:nor_result}
\end{figure}

We simulate the range of SCO rates that contain the spread of infection even if some contacts do not receive tests and quarantines. Figure~\ref{fig:nor_result} shows the results of the spread of infection according to the SCO rate in virtual Seoul. The mean cumulative number of infections increases as the SCO rate rises (Figure~\ref{fig:nor_result}(a)). Figure~\ref{fig:nor_result}(b) shows that the mean time of the epidemic peak is progressively delayed. In contrast to the IO scenario, which shows a sharp transition, the CO scenario exhibits a continuous delay as the omission rate increases (Figure~\ref{fig:nor_result}(c)). This suggests that missing contacts does not immediately trigger an explosive outbreak but instead allows infection to spread gradually through the population. As a result, transmission persists without immediate stochastic extinction while avoiding a rapid outbreak. The mean diameter of the directed transmission network does not increase beyond a certain size. In Figure~\ref{fig:nor_result}(d), the out-degree distribution retains a fat-tailed form. Because not all missed contacts result in secondary infections, increases in the SCO rate lead to less pronounced acceleration of transmission compared with the IO scenario.

\begin{figure}
\centering
\includegraphics[width=0.6\textwidth]{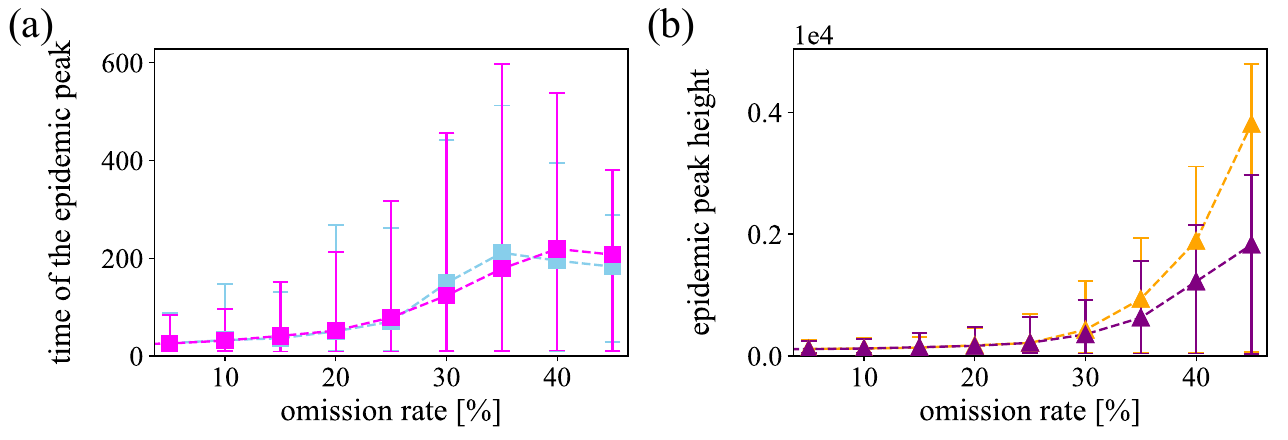}
\caption{Comparison of the two contact-omission scenarios in virtual Seoul. Results are shown for $E_0 = 20$ with a 95\% CI: (a) the mean time of the epidemic peak and (b) the mean epidemic peak height against the CO rates. In (a), magenta denotes the SCO scenario and sky-blue denotes the UCO scenario; in (b), purple denotes the SCO scenario and orange denotes the UCO scenario.}
\label{fig:nor_diff}
\end{figure}

In addition, we simulate the UCO scenario that contacts are missed equally in households, school classrooms, workplaces, and friends' gatherings (the omission rate is 50\% in the local community). In Figure~\ref{fig:nor_diff}, the top panel shows the results for the SCO scenario (omissions in the friends' gathering and local community networks), whereas the bottom panel illustrates the UCO scenario (omissions across all five network layers). Both CO scenarios exhibit similar overall trends. As the CO rate increases, the mean time of the epidemic peak is delayed, and, the mean epidemic peak height becomes larger. The overall scale of transmission remains smaller than that observed under the IO scenario. These findings suggest that accurately reconstructing the movement trajectories of confirmed cases is more crucial than notifying every contact.

We also repeat the same simulations by doubling the number of initially $E$ agents ($E_0 = 40$). The initial size of the $E$ population has minimal influence on the overall infection dynamics (see, SI). Although SCO and UCO have a less pronounced impact than IO, higher omission rates still accelerate transmission and reduce the overall effectiveness of manual CT in containing outbreaks, highlighting the importance of accurate trajectory reconstruction for sustained epidemic control.

\section*{Discussion}
Using a high-resolution agent-based model (ABM) with synthetic populations for Seoul and Busan, we quantified how information omissions in manual contact tracing (CT) affect its ability to contain an emerging infectious disease during the early phase of an outbreak. We distinguished two types of omission: failure to reconstruct the movement trajectories of confirmed cases (infector-omission, IO) and failure to notify identified contacts for testing and quarantine (contact-omission, CO). Our simulations indicate that manual CT can slow early spread even when part of this information is missing, but only up to a city-specific threshold. In virtual Seoul, the IO rate could rise to about 4\% before CT effectiveness collapsed, whereas in virtual Busan the corresponding threshold was higher, at about 10\%. Beyond these levels, the cumulative number of infections increased rapidly, while the diameter of the transmission network no longer grew but saturated at a finite value. Across both cities, IO had a consistently greater impact on containment than CO. We emphasize that these threshold estimates apply specifically to the early-outbreak, near-instantaneous-tracing regime considered in the main analysis; under sustained tracing delays (Sec.~S5.1 of SI), the outbreak saturates regardless of the omission rate, and a fixed numerical threshold no longer captures the boundary of effective containment.

Busan has a lower population density and an older demographic structure than Seoul. Its higher tolerance to omission likely reflects differences in population size, contact density, and commuting intensity, consistent with evidence that inter-city interactions and population structure shape how epidemics scale across urban areas~\cite{loureiro2025impact}. These results suggest that CT effectiveness depends on city-specific demographic and social characteristics, and that tracing strategies should be tailored to regional population and mobility structures~\cite{zhang2023evaluating}. 

At low omission rates, the main effect of IO was to delay the epidemic peak rather than to prevent the outbreak. This should be interpreted with caution, since a delayed peak can mask the fact that the system is approaching the threshold beyond which transmission becomes self-sustaining. CO failures, by contrast, produced a more gradual shift in epidemic timing. A plausible mechanistic explanation is that missing an infector (IO) leaves an entire downstream transmission chain untraced, preserving much of the connectivity of the transmission network, whereas missing a single contact (CO) removes only one link. Under this interpretation, trajectory reconstruction acts as a more upstream intervention and may therefore be more important for keeping the effective reproduction number below unity.

A further contribution of the study is the topological analysis of the transmission network through its diameter. As the IO rate increased, the diameter grew, indicating longer transmission chains, and then saturated at a finite value beyond the threshold, so that additional spread was reflected in the breadth rather than the depth of the transmission tree even as the cumulative number of infections continued to rise. Whereas epidemic size reflects the immediate disease burden, the diameter captures how many successive transmission events the pathogen passes through. We note that our model does not simulate viral evolution. Nonetheless, the deeper transmission trees associated with higher IO would, in principle, offer more opportunities for mutation accumulation, a possibility that warrants explicit evolutionary modeling in future work.

Translating these thresholds into practice requires operational proxies that public health authorities can monitor in near real time. For the IO rate, several routinely collected indicators could serve as proxies: the proportion of confirmed cases for which an epidemiological investigation is not initiated or completed, the delay between case confirmation and the start of the investigation, and the caseload per investigator. The first directly approximates the fraction of cases whose trajectories go unreconstructed, while the latter two act as leading indicators, since longer delays and heavier caseloads tend to precede a rise in untraced cases. We quantify the epidemic impact of such a confirmation-to-tracing delay in Sec.~S5.1 of SI. For the CO rate, the contact follow-up completion rate offers a natural proxy, defined as the proportion of the contacts named by a confirmed case who are successfully reached for testing and quarantine. During an investigation, a confirmed case is asked to recall whom they have recently met. Among the contacts they name, some cannot be placed under quarantine because their identity cannot be established or because they have already left the country, and the share of such unreachable contacts directly reflects the CO rate. Because these indicators are, in principle, already captured in routine investigation records, authorities could track them against the city-specific thresholds identified here and treat a sustained approach toward those levels as an early warning that CT is nearing the point of diminishing effectiveness. We emphasize, however, that these proxies are approximate and would need calibration against local investigation practices before operational use.

Several limitations of this study should be acknowledged. First, our simulations were not based on actual CT logs or epidemiological investigation data, but were constructed from census microdata, social contact survey data, and other statistical sources to create synthetic yet realistic contact networks~\cite{kerr2021covasim}. We were therefore unable to quantitatively validate the realism of the CT processes reproduced by the model. Second, the speed of CT is itself decisive. Our main analysis assumed instantaneous tracing. In Sec.~S5.1 of SI, we relax this assumption and show that even a uniform two-day delay between case confirmation and contact quarantine is, by itself, sufficient to drive a large-scale outbreak across all omission scenarios. We did not, however, model an initial preparation phase before the first cases are traced, nor delays that grow dynamically with investigator caseload as the epidemic progresses. Capturing such time-varying delays remains for future work. Third, dynamic changes in individual behavior or government policy responses over time were not incorporated. Fourth, because each omission rate requires an ensemble of large-scale stochastic simulations run until the epidemic dies out, we evaluated outcomes at a set of representative omission rates rather than performing formal statistical tests across the full range. The sampled intervals were nonetheless chosen finely enough to capture the overall trends and the threshold behavior. Together, these assumptions mean that the model does not fully capture how omissions arise in real-world CT operations.

Nevertheless, we confirmed that our main qualitative conclusions persist when infector- and contact-omission occur simultaneously (Sec.~S5.2 of SI) and across alternative values of the self-testing probability $P_t$ (Sec.~S5.3 of SI), although the absolute epidemic size is itself sensitive to $P_t$. To address these limitations, future work should incorporate actual CT logs or investigation records for model validation, calibrate the operational proxies proposed above against real investigation data, and explore policy scenarios that include adaptive behavioral and governmental responses.

Despite these limitations, our study has several notable strengths. Few quantitative assessments of CT have distinguished infector- and contact-omission as mechanistically distinct processes or linked them to the structure of the transmission network, with most treating information loss as a single aggregate quantity. Our study does both, building on a realistic multilayer contact network for two distinct metropolitan areas to provide quantitative estimates of the impact and threshold levels of CT information loss through city-scale, high-resolution simulations. These findings offer a practical starting point for designing CT strategies under constrained resources, and simulation-based analyses such as this can support public health decision-making by helping to define acceptable levels of information loss and to prioritize response actions. Future research may extend the model to incorporate vaccination coverage, the transmissibility of emerging variants~\cite{wang2022modeling}, and digital tracing technologies, enabling more refined and policy-relevant epidemic assessments.

\section*{Data availability}
The simulation codes supporting the findings of this study are available at \url{https://github.com/MkChae/ABM_CT}. All input data required to reproduce the simulations, including synthetic population data, are available at \url{https://doi.org/10.6084/m9.figshare.31142995}.

\section*{Funding}
This work was supported by the National Institute for Mathematical Sciences (NIMS) under Grant No.~NIMS-B26730000 (W.-S.S.) and by the National Research Foundation of Korea (NRF) under Grant No.~RS-2026-25468383 (S.H.L.).

\bibliography{CT_ABM_ref_revised}




\section*{Author contributions statement}
M.-K.C. and W.-S.S. designed the study. M.-K.C. developed the model and conducted the simulations. M.-K.C. and S.H.L. analyzed the results. All authors conducted the literature review, wrote the manuscript, and reviewed the manuscript.

\section*{Competing interests}
The authors declare no competing interests.




\clearpage

\end{document}